\documentclass[prd,twocolumn,showpacs,preprintnumbers,amsmath,amssymb]{revtex4-2}
\usepackage{amsfonts,amssymb,amsmath,mathrsfs}
\usepackage{url}
\usepackage{graphicx}
\newtheorem{thrm}{Theorem}
\newtheorem{lem}[thrm]{Lemma}

\newtheorem{remark}[thrm]{Remark}
\begin{document}
\title[Conformal Killing gravity]{A note on Harada's Conformal Killing gravity} 

\author{Carlo Alberto Mantica}
\author{Luca Guido Molinari} 
\affiliation{Physics Department Aldo Pontremoli,
Universit\`a degli Studi di Milano and I.N.F.N. sezione di Milano,
Via Celoria 16, 20133 Milano, Italy.}
\email{carlo.mantica@mi.infn.it, luca.molinari@unimi.it}

\begin{abstract} We show that ``Gravity at cosmological distances: explaining the accelerating expansion without dark energy'' recently proposed by J. Harada \cite{Harada23} is equivalent to the Einstein 
equation extended by the presence of an arbitrary conformal Killing tensor. This turns Harada's equations of third order in the derivatives of the metric tensor to second order, and offers a strategy of solution that covariantly shortcuts Harada's derivation and obtains both modified Friedmann equations. Another illustration is presented for the case of flat space and constant curvature. 
\end{abstract}
\date{10 oct 2023}


\maketitle
\section{Introduction}
In the recent paper \cite{Harada23} in this journal, Junpei Harada posed three theoretical criteria for gravitational theories:
1) the cosmological constant $\Lambda $ is obtained as a constant of integration;
2) the stress-energy conservation law $\nabla_j T^{jk}=0$ is derived as a consequence of the gravitational field equations, rather than being assumed;
3) a conformally flat metric is not necessarily a solution in the vacuum.\\
Based on these criteria, he proposed the new gravitational equations:
\begin{align}
H_{jkl} =& 8\pi G \,T_{jkl} \label{1.1} \\
H_{jkl}=& \nabla_j R_{kl} +  \nabla_k R_{lj}  + \nabla_l R_{jk} \nonumber\\
& -\tfrac{1}{3} (g_{kl} \nabla_j R + g_{lj} \nabla_k R + g_{jk} \nabla_l R )  \nonumber\\
T_{jkl}=& \nabla_j T_{kl} +  \nabla_k T_{lj}  + \nabla_l T_{jk} \nonumber \\
& -\tfrac{1}{6} (g_{kl} \nabla_j T + g_{lj} \nabla_k T + g_{jk} \nabla_l T) \nonumber
\end{align}
$R_{jk}$ is the Ricci tensor with trace $R$, $T_{kl}$ is the stress-energy tensor with trace $T$. The Bianchi identity 
$\nabla_j R^j{}_k = \frac{1}{2}\nabla_k R$ implies $\nabla_j T^j{}_k=0$. 
Solutions of the Einstein equation are solutions of the new theory.

Harada solved them in spherical vacuo and in cosmology. In the first case,
$H_{jkl}=0$, he searched for a static spherically symmetric solution $ds^2 = - e^\nu dt^2 +  e^{-\nu} dr^2 + r^2 (d\theta^2 +\sin^2 d\phi^2)$. He obtained $ e^\nu = 1- \frac{2M}{r} -\frac{\Lambda}{3} r^2 - \frac{\lambda}{5} r^4 $ i.e.
the Schwarzschild term, a de-Sitter term with cosmological constant, and a new term that dominates at large distances. 
The most general static spherical solution was shortly after obtained by Alan Barnes \cite{Barnes}.

In cosmology, Harada obtained an equation of motion for the scale factor of a RW spacetime  describing the transition from decelerating to accelerating expansion in a matter-dominated universe, without dark matter and  cosmological constant. The analysis is deepened in his subsequent draft \cite{Harada23b} where radiation is also accounted for.

Eq. \eqref{1.1} contains third order derivatives of the metric. 
This feature was also present in his Cotton gravity theory \cite{Harada21}. However, we showed in \cite{Codazzi} that the Cotton theory is equivalent to an extended Einstein gravity where the stress-energy tensor in the Einstein equation is modified by an arbitrary Codazzi tensor. This amounts to an integration of the Cotton gravity equation, that reduces to second order with the appearance of a supplemental term. \\
In this note we show that eq.\eqref{1.1} can be recast as an Einstein 
equation with the stress-energy tensor modified by an arbitrary divergence-free conformal Killing tensor.\\
We then apply a strategy to solve the modified Einstein equation in a Robertson-Walker spacetime, 
and obtain the 
modified Friedmann equations. We shortly revisit the case $k=0$, $\Lambda =0$ and zero pressure studied by Harada, 
and discuss another soluble case: $k=0$, $\Lambda =0$ and $\nabla_k R=0$.

\section{Conformal Killing gravity}
Conformal Killing tensors are well known in differential geometry \cite{Rani03,Coll06,Sharma10,Kobialko22}. They are symmetric tensors $K_{ij}$
characterized by the condition
\begin{align}
\nabla_j K_{kl} +  \nabla_k K_{lj}  + \nabla_l K_{jk} = \eta_j g_{kl} + \eta_k g_{lj}+ \eta_l g_{jk}  \label{1.4}
\end{align}
where the non-vanishing 1-form $\eta_i$ is the {\em associated conformal vector}. Killing tensors are 
recovered whenever $\eta_i=0$. If $\eta_j = \nabla_j \phi$ then $K_{ij}$ is a {\em gradient conformal Killing tensor}.\\
The contraction of \eqref{1.4} with $g^{kl}$ shows that $\nabla_j K^{jl}=0$ if and only if $\eta_j = \frac{1}{6} \nabla_j K$, where $K$ is the trace.

The gravity theory proposed by Harada is deeply connected with conformal Killing
tensors. Eq.\eqref{1.1} is easily rewritten (hereafter we set $8\pi G=1$) as
\begin{align}
& \nabla_j (R_{kl}-\frac{R}{2} g_{kl}  -T_{kl}) \\
&+  \nabla_k (R_{lj} -\frac{R}{2} g_{lj} - T_{lj})   +  \nabla_l (R_{jk}-\frac{R}{2} g_{jk} - T_{jk}) \nonumber \\
&= - \frac{1}{6}\left[g_{kl} \nabla_j (R+T) + g_{lj} \nabla_k (R+T)  + g_{jk} \nabla_l (R+T)\right ]. \nonumber
\end{align}
The equation defines $K_{jk} = R_{jk}-\frac{R}{2} g_{jk} - T_{jk}$, with trace $K=-R-T$, as a gradient conformal Killing tensor with associated conformal vector $\eta_j = \frac{1}{6}\nabla_j K$.\\
 Eq.\eqref{1.1} is thus equivalent to
\begin{align}
&R_{jk} -\frac{R}{2} g_{jk} = T_{jk} + K_{jk} \label{1.7}\\
&\nabla_j K_{kl} +  \nabla_k K_{lj} +  \nabla_l K_{jk} \label{1.7b}\\
&=  \frac{1}{6}( g_{kl}\nabla_j K  + g_{lj} \nabla_k K + g_{jk} \nabla_l K  ). \nonumber
\end{align}
%
\textit{Harada’s gravitational equation \eqref{1.1} is equivalent to the Einstein equation in which the stress-energy tensor is modified by a divergence-free conformal Killing tensor.}\\
$\bullet $ The property $\nabla_j K^{jl}=0$ ensures that $\nabla_j T^{jl}=0 $.\\
$\bullet $ The third order character of eq.\eqref{1.1} reduces to second order in \eqref{1.7} with the appearance of a conformal Killing tensor in the Einstein equation. 

We may name the new theory \textsf{Conformal Killing gravity}.
Being appropriate, the name was adopted by Harada in his subsequent paper on cosmology \cite{Harada23b}.

Two interesting remarks:

1) The vacuum equation of Conformal Killing gravity is $H_{jkl}=0$ i.e. the Ricci tensor is itself conformal Killing. An example are the Sinyukov spacetimes, defined by $\nabla_j R_{kl} = \frac{1}{18}(4g_{kl}\nabla_j R + g_{jl}\nabla_k R + g_{jk} \nabla_l R)$. They have zero Cotton tensor, and were studied by Formella \cite{Formella95}. They are one of the seven symmetry classes in Gray's decomposition of $\nabla_j R_{kl}$ \cite{GRAY}.

2) Theorem 1 in \cite{Rani03} gives the explicit construction of conformal Killing tensors from conformal Killing vectors, 
$\nabla_i X_j +\nabla_j X_i = 2\psi g_{ij} $, where the function $\psi $ is named conformal factor. 
With a single vector, the tensor is 
\begin{align}
K_{ij} = \alpha X_i X_j +\beta g_{ij} \label{KXX}
\end{align}
where $\alpha $ is a constant and $\beta $ is a function. The additional condition 
$\nabla_i K^i{}_j =0$ constrains $\beta $.\\
Such vectors frequently characterize spacetimes, with a wide range of possibilities. For example Bang Yen Chen characterized generalized Robertson-Walker spacetimes through a time-like vector 
$\nabla_i X_j = \psi g_{ij}$  \cite{BYChen}. Ramos et al. characterized the ample class of doubly warped spacetimes 
by conformal Killing vectors \cite{Ramos,DWSP}.

\section{Conformal Killing tensor for cosmology}
To find solutions of eq.\eqref{1.7} we apply the same strategy in our study of Cotton gravity \cite{Codazzi}. We  first fix a physically appropriate form of the conformal Killing tensor \eqref{1.7b}. This fixes the spacetime, where the form of the Ricci tensor is then determined. Finally, the stress-energy tensor is obtained by eq.\eqref{1.7}.\\
In doing so for a cosmological solution, we obtain in natural and easy fashion the results by Harada.

The cosmological principle stages any gravitational theory in a Robertson-Walker spacetime, which is quasi-Einstein
i.e. the Ricci tensor has the perfect fluid form. We then consider a tensor of the same form, and pretend 
that it is a conformal Killing tensor,
\begin{align*}
K_{ij} =  \frac{K-\lambda}{3} g_{ij} + \frac{K-4\lambda}{3} u_i u_j 
\end{align*}
with $u^ju_j=-1$, $\nabla_j u_k = \nabla_k u_j$. $K$ is the trace and $\lambda $ is an eigenvalue: $K_{ij} u^j = \lambda u_i$. We also require that the associated conformal vector is $\eta_j = \frac{1}{6}\nabla_j K$.

There are two useful facts proven in the survey \cite{Survey}. 
The first one is:
\begin{lem}[Lemma 4.2 in \cite{Survey}]
Let $K_{ij}$ a conformal Killing tensor and $K_{ij}u^j = \lambda u_i$, $u^j u_j =\pm 1$. If $u^k\nabla_k u_j =0$ then the associated conformal vector is $\eta_j = \nabla_j \lambda $. 
\end{lem}
\noindent
The Lemma implies the important relation $\frac{1}{6}\nabla_j K = \nabla_j \lambda $ i.e. 
$K= 6\lambda + 2\Lambda $, where $\Lambda $ is an integration constant. The updated conformal Killing tensor 
\begin{equation}
K_{ij} = \frac{5\lambda + 2\Lambda}{3} g_{ij} + \frac{2\lambda + 2\Lambda}{3} u_i u_j  \label{1.9}
\end{equation}
by construction fulfils \eqref{1.7b} and $\nabla_j K^{jl}=0$. 

The other useful statement characterises the spacetime that hosts this conformal Killing tensor.\\
Let us recall that a generalized Robertson-Walker (GRW) spacetime is characterized by the metric 
$$ds^2 = -dt^2 + a^2(t) g^\star_{\rho\nu} ({\bf x}) dx^\rho dx^\nu$$
where $a(t)$ is the scale function, $g^\star_{\rho\nu}$ is the metric tensor of the Riemannian space sub-manifold. 
An equivalent description is the existence of a 
velocity field $u_j$, $u^j u_j=-1$ that is shear, rotation and acceleration free:
\begin{align}
 \nabla_i u_j = H (u_i u_j + g_{ij} ) \label{VEL}
 \end{align}
with $H$ a scalar function such that $\nabla_j H = - \dot H u_j$. In coordinates $(t,{\bf x})$, $H$ is only a function of time: $H(t)=\dot a/a $, and has the same role as the Hubble parameter in Robertson-Walker (RW) spacetimes. A RW spacetime
is the special case where the Weyl tensor is zero.\\
Note that the velocity is closed, so that the  acceleration $u^k\nabla_k u_j$ is zero.
\begin{thrm} [Theorem 4.3. in \cite{Survey}]
A Lorentzian manifold is a GRW spacetime if and only if there is a conformal Killing 
tensor of the form $K_{ij} = A g_{ij} +B u_i u_j$, where $A$, $B$ are scalar fields, $B\neq 0$, 
$u_j u ^j = -1$ and $\nabla_j u_k= \nabla_k u_j$.\\
The associated conformal vector is $\eta_j = \nabla_j A + \dot B u_j$. \\
The velocity field satisfies $\nabla_j u_k = \frac{\dot B}{2B} (u_j u_k + g_{jk})$, and $\nabla_j B = -\dot B u_j$.  
\end{thrm}

According to the theorem, $K_{ij}$ in \eqref{1.9} is a conformal Killing tensor if and only if the spacetime is a GRW. In this circumstance the Hubble parameter is
$$ H = \frac{\dot a}{a} =\frac{ \dot \lambda}{2\lambda + 2\Lambda } $$
and $\nabla_j \lambda = -  \dot \lambda u_j$. The conformal vector is $
\eta_j = \frac{1}{3} \nabla_j (5\lambda) + \frac{1}{3} (2\dot\lambda) u_j = \nabla_j \lambda $ as expected. 
The equation for $H$ has the solution ($C$ is a constant):
\begin{align}
2\lambda + 2\Lambda = C \; a^2  \label{lambdaLambda}
\end{align}

In a GRW spacetime the Ricci tensor is (\cite{Survey}, eq.17)
\begin{align}
R_{kl} = \frac{R-4\xi}{3} u_k u_l + \frac{R-\xi}{3} g_{kl}  - 2 C_{jklm} u^j u^m \label{1.14}
\end{align}
The last term is the electric component of the Weyl tensor, that vanishes in a RW spacetime. $\xi = 3(\dot H + H^2) = 3 \ddot a / a$ is the eigenvalue ($R_{ij}u^j =\xi u_i$)
and the curvature scalar is
\begin{align}
 R = \frac{R^\star}{a^2} + 6 \frac{\dot a^2}{a^2} + 6\frac{\ddot a}{a} \label{Rscalar}
 \end{align}
where $R^\star $ is the curvature  scalar of the space sub-manifold.
The Ricci and the conformal Killing tensors in eq.\eqref{1.7} give the stress-energy tensor:
\begin{align}
&T_{jk} = R_{jk} -\frac{R}{2}g_{jk} - K_{jk} \\
&= -\frac{1}{3} ( \frac{R}{2} +\xi + 5\lambda +  2\Lambda  ) g_{jk}  + \frac{1}{3} \left ( R-4\xi - 2\lambda - 2\Lambda \right ) u_j u_k \nonumber
\end{align}
The tensor describes a ``cosmological'' perfect fluid corrected by the geometry of the new theory. Comparison
with the fluid parameterization $T_{jk} = p g_{jk} +(p+\rho) u_j u_k$ gives the first and second modified Friedmann equations, where we reintroduce Newton's constant:
\begin{align}
8\pi G p =&-\frac{R}{6} -\frac{\xi}{3} -\frac{5}{3}\lambda -\frac{2\Lambda}{3} \nonumber\\
=& -\frac{R^\star}{6a^2} - \frac{\dot a^2}{a^2} - 2\frac{\ddot a}{a} + \Lambda   - \frac{5}{6}C a^2  \label{FF1} \\
8\pi G \rho =&\frac{R}{2} - \xi + \lambda  \nonumber\\
=& \frac{R^\star}{2a^2} +3 \frac{\dot a^2}{a^2} -  \Lambda + \frac{1}{2}C a^2  \label{FF2} 
\end{align}
With $C=0$ they are the standard Friedmann equations with cosmological constant $\Lambda $, that here entered as
an integration constant. \\
Elimination of $C$ gives eq.32 by Harada \cite{Harada23}
\begin{align}
4\pi G \frac{5\rho + 3p}{3} = \frac{R^\star}{3a^2} + 2 \frac{\dot a^2}{a^2} -  \frac{\ddot a}{a} -\frac{\Lambda}{3} \label{HH}
\end{align}
with the identification $R^\star /6 =k$.\\
With an equation of state $p=w\rho$, the equation of continuity gives $\rho (a) =\rho_0 (a_0/a)^{3(w+1)}$. With $a(-t_0)=0$, $a(0)=a_0$,  $D\equiv 8\pi G \rho_0 a_0^3$, eq.\eqref{FF2}
is formally integrated:
\begin{align}
t+t_0 = \int_0^a dx \left[\frac{D a_0^{3w}}{x^{3w+1}} - \frac{R^\star}{6}+\frac{\Lambda}{3} x^2 -\frac{1}{6}Cx^4
\right]^{-\frac{1}{2} } \label{14}
\end{align}

\begin{remark}
The conformal Killing tensor \eqref{1.9} may be obtained through eq.\eqref{KXX} from the time-like conformal Killing vector 
$X_i = F u_i$ (\cite{El2017}, thrm 1). 
The equation implies $\psi = FH=\dot F$ i.e. $F= c a(t)$. In $K_{ij} = (\alpha F^2) u_i u_j + \beta g_{ij} $ the condition of
zero-divergence poses $\beta = \frac{5}{2} \alpha F^2 - \Lambda $:
$$K_{ij} = (\alpha F^2) u_i u_j + (\frac{5}{2} \alpha F^2 - \Lambda) g_{ij} $$ 
that is \eqref{1.9} with $K=9\alpha F^2 -4\Lambda $ and $\lambda = \frac{3}{2}\alpha F^2 -\Lambda$.

In a GRW space-time the torse-forming vector field $u_i$ with $\dot H\neq 0$  is unique (\cite{TACCHINI} thrm.B). 
Then the time-like conformal Killing vector is unique.
\end{remark}

\subsection{Model $\mathbf{p=0}$ (matter dominated universe)} There is consensus that the large scale universe is characterized by a spatial curvature $R^\star$ very near zero \cite{Wang20,Melia22}.
Harada studied the continuity equation and \eqref{HH} in the case $p=0$ (dust matter M), $R^\star =0$ and $\Lambda =0$.

We rewrite eq.\eqref{FF1} for $p=0$ without restrictions:
$$0=2\frac{d}{dt} (\dot a\sqrt a) + \frac{5}{6}C a^{7/2} - \Lambda a^{3/2} +  \frac{R^\star}{6\sqrt a}$$ 
Multiplication by $\dot a \sqrt a$ gives a total derivative. Then:
\begin{align}
H^2(a) = -\frac{C}{6} a^2 + \frac{\Lambda}{3}  - \frac{R^\star}{6a^2} + \frac{D}{a^3}
\end{align} 
where $D$ is an integration constant. With $\Lambda=R^\star=0$ it is eq.41 in \cite{Harada23}. Eq.\eqref{FF1} becomes:
$   \frac{\ddot a}{a} = \frac{\Lambda}{3}   - \frac{1}{3}C a^2 - \frac{D}{2a^3}$.
The results for $H^2$ and $\ddot a/a$ in eq.\eqref{FF2} give, as expected:
$\frac{8}{3}\pi G \rho = D/a^3$. With present time values $a_0$, $H_0$, $\rho_0$ it is
$ D= \frac{8}{3}\pi G \rho_0 a_0^3 $.

The integral \eqref{14} with $w=0$, $R^\star=0$ and $\Lambda =0$ gives Harada's result:
\begin{align}
t+t_0 = \frac{2}{3\sqrt D}a^{3/2} {}_2F_1\left (\frac{1}{2}, \frac{3}{10}; \frac{13}{10}; \frac{C}{6D} a^5 \right ). 
\end{align}
With the (standard) definitions in \cite{Harada23} the constants are:
$$ \frac{C}{6}=\frac{H_0^2}{a_0^2} (\Omega_M -1), \qquad  \frac{C}{6D} = 
\frac{1}{a_0^5}\frac{\Omega_M -1}{\Omega_M} $$

In \cite{Harada23b} Harada solved the modified Friedmann equations \eqref{1.7}\eqref{1.7b} by also including radiation.
He finds that the far future evolution is dominated by the conformal Killing term, with a phantom energy $w=-5/3$.

\subsection{Model $\mathbf{R^\star =0}$ and ${\mathbf R}$ constant}
In \cite{COSMOLOGY} we studied a RW spacetime with $R^\star=0 $ and $R$ constant. Eq.\eqref{Rscalar} is $\frac{d^2}{dt^2}a^2  = (R/3) a^2$. With the initial condition $a(0)=0$, the solution is
$$ a^2(t) = A^2 \sinh \theta , \qquad \theta=t \sqrt{\frac{R}{3}}$$
where $A$ is a constant and the constant curvature $R$ is a time-scale. Then
\begin{align}
 H(t) = \frac{\dot a}{a} = \frac{1}{2} \sqrt{\frac{R}{3}} {\rm coth} \theta \label{HUBBLE}
 \end{align}
%
This time-evolution is fixed by the RW geometry with $R^\star =0$ and $\nabla_k R=0$.\\
The Friedmann equations yield the pressure $p(t)$ and energy density $\rho (t)$ of the cosmological fluid. The outcome with $\Lambda =0$, is an equation of state $p=w\rho $ with $w\to 1/3$ in the early universe ($t\sqrt R\ll 1$) and $w=-1$ in the late universe ($t\sqrt R \gg 1$), i.e. a transition from a radiation era to a dark energy era in a time-lag that, with Planck's data, is compatible with the age of the universe.

Now we study the same geometric conditions (yielding same function $a^2(t)$) in conformal Killing gravity.\\  
Eq.\eqref{lambdaLambda} is
$2\lambda (t)= Ca^2(t)$. The Friedmann equations are modified by the conformal Killing tensor by terms proportional to $C$. They control the late evolution:
\begin{align}
& 8\pi G \rho (t) =\frac{R}{4} \coth^2 \theta   +\frac{CA^2}{2} \sinh \theta  \\
&8\pi G(\rho (t) -3p(t)) = R +3\, CA^2 \sinh \theta 
\end{align}
The equations describe a radiation dominated universe $p\approx\frac{1}{3}\rho $ for $t\sqrt R \ll 1$, and a phantom energy 
era \cite{Caldwell03} $p\approx -\frac{5}{3}\rho $ for $t\sqrt R \gg 1$, as in Harada's model \cite{Harada23b}.\\

The detailed evolution is obtained by assuming that the universe is filled with radiation $R$, matter
$M$ and a dark fluid $D$ with unknown function $D(x)$, $D(1)=1$:
$$\frac{\rho_R (t)}{\rho_{R0}}= \left [\frac{a}{a_0}\right ]^{-4},\; \frac{\rho_M(t)}{\rho_{M0}}= \left[\frac{a}{a_0}\right ]^{-3}, \;
\frac{\rho_D (t)}{\rho_{D0}}= D \left (\frac{a(t)}{a_0}\right ) $$
The Friedmann equation \eqref{FF2} with $\rho = \rho_R + \rho_M + \rho_D$ and 
$R^\star =0$ is:
\begin{align}
\frac{H^2}{H_0^2} = \Omega_{R}  \left [\frac{a(t)}{a_0}\right ]^{-4}+\Omega_M  \left [\frac{a(t)}{a_0}\right ]^{-3} \nonumber \\ 
+ \Omega_D D(\frac{a(t)}{a_0})
 - \frac{Ca_0^2}{6H_0^2} \left [\frac{a(t)}{a_0}\right ]^2 \label{HH0}
\end{align}
where $\Omega_R= (8\pi G /3H_0^2) \rho_{R0}$ and similar for $\Omega_M$ and $\Omega_D$.
The ratio $H/H_0$ is fixed by the geometry, eq.\eqref{HUBBLE}. Then we obtain $D$ (here written as a function of $\theta $):
\begin{align*}
\Omega_D  D(\theta) =& {\rm th}^2\theta_0 \left [1+\frac{1-\alpha}{{\rm sinh}^2\theta} -  \frac{\alpha \Omega_M}{\Omega_R^{3/4} (\alpha -\Omega_R)^{1/4} } \frac{1}{{\rm sinh}^{3/2}\theta } \right ]\\
& + \frac{Ca_0^2}{6H_0^2} \frac{{\rm sinh}\theta}{\rm{sinh}\theta_0} 
\end{align*}
where we defined $\alpha = \Omega_R {\rm cosh}^2\theta_0$ (\cite{COSMOLOGY}). The above equation extends eq.24 in
\cite{COSMOLOGY}.\\
 For $a=a_0$ we read in \eqref{HH0} the present-time balance of the various components: 
\begin{align*}
 \Omega_R + \Omega_M + \left[\Omega_D - \frac{Ca_0^2}{6H_0^2}\right ] =1 
 \end{align*}
In parenthesis is the dark term of the theory, with an explicit contribution from the conformal Killing term ($C\neq 0$)
and a remaining part that emerges from the extended Einstein equation. 

In this model we used a non-standard approach by fixing a-priori the geometry with the reasonable constraint $R^\star =0$ and the simple
condition $\nabla_k R=0$. 
The standard approach to cosmology in GR is to give a specific equation of state for the dark sector, which 
together with matter and radiation determine the ratio $H(z)/H_0$. A general discussion, where the dark sector is described by an equation of
state with $w$ a function of the redshift $z$ is found in \cite{Bargiacchi}.
\vfill

\subsection*{Data availability}
Data sharing is not applicable to this article as no datasets were generated or analyzed during the current study.


\begin{thebibliography}{99}
%
\bibitem{Bargiacchi}
G.~Bargiacchi, M.~Benetti, S.~Capozziello, E.~Lusso, G.~Risaliti and M. Signorini, 
\textit{Quasar cosmology: dark energy evolution and spatial curvature}, MNRAS {\bf 515} (2) (2022) 1795--1806.
%
\bibitem{Barnes}
A.~Barnes, \textit{Vacuum static spherically symmetric spacetimes in Harada's theory}, (2023) arXiv:2309.05336 [gr-qc]
%
\bibitem{Caldwell03}
R.~R.~Caldwell, M.~Kamionkowski and N.~N.~Weinberg, \textit{Phantom energy and cosmic doomsday}, 
Phys. Rev. Lett. {\bf 91} 071301, (2003)
%
\bibitem{BYChen}
B.-Y.~Chen, \textit{A simple characterization for generalized Robertson-Walker spacetimes}, Gen. Relativ. Gravit. {\bf 46}, 1833 (2014). 
%
\bibitem{Coll06}
B.~Coll, J.~J.~Ferrando, J.~A.~S\'aez, \textit{On the geometry of Killing and conformal tensors}, J. Math. Phys. {\bf 47}, 062503 (2006). 
%
\bibitem{El2017}
H.~K.~El-Sayed, S.~Shenawy and N.~Syied, \textit{On symmetries of generalized Robertson-Walker spacetimes and applications}, J. Dynamical Systems and geometric theories {\bf 15} (1) (2017) 51--69.
%
\bibitem{Formella95}
S.~Formella, \textit{On some class of nearly conformally symmetric manifolds}, Colloq. Math. {\bf 68} (1) 149--164 (1995). 
%
\bibitem{Harada21}
J.~Harada, \textit{Emergence of the Cotton tensor for describing gravity}, Phys. Rev. D  {\bf 103}, L121502 (2021). 
%
\bibitem{Harada23}
J.~Harada, \textit{Gravity at cosmological distances: Explaining the accelerating expansion without dark energy}, 
Phys. Rev. D {\bf 108}, 044031 (2023).
%
\bibitem{Harada23b}
J.~Harada, {\em Dark energy in conformal Killing gravity}, arXiv:2308.07634 [gr-qc].
%
\bibitem{Kobialko22}
K.~Kobialko, I.~Bogush, and D.~Gal'tsov, \textit{Slice-reducible conformal Killing tensors}, Phys. Rev. D {\bf 106}, 024006 (2022). 
%
\bibitem{Survey}
C.~A.~Mantica and L.~G.~Molinari, \textit{Generalized Robertson-Walker spacetimes, a survey}, Int. J. Geom. 
Meth. Mod. Phys. {\bf 14} n.3, 1730001 (2017). 
%
\bibitem{COSMOLOGY}
C.~A.~Mantica and L.~G.~Molinari, \textit{w=1/3 to w=-1 evolution in a Robertson-Walker spacetime with constant scalar curvature}, Int. J. Geom. Meth. Mod. Phys. {\bf 16}, 1950061 (2019).
%
\bibitem{GRAY}
C.~A.~Mantica, L.~G.~Molinari, Y.~J.~Suh and S.~Shenawy, 
\textit{Perfect-fluid, generalized Robertson-Walker spacetimes, and Gray’s decomposition},
J. Math. Phys. {\bf 60}, 052506 (2019).
%
\bibitem{DWSP}
C.~A.~Mantica and L.~G.~Molinari, \textit{Spherical doubly warped spacetimes for radiating stars and cosmology}, 
Gen. Relativ. Gravit. {\bf 54}:98 (2022). 
%
\bibitem{Codazzi}
 C.~A.~Mantica and L.~G.~Molinari, \textit{Codazzi tensors and their spacetimes, and Cotton gravity}, 
 Gen. Relativ. Gravit. {\bf 55}:62 (2023).
%
\bibitem{TACCHINI}
L.~G.~Molinari, A.~Tacchini and C.~A.~Mantica, \textit{On the uniqueness of a shear-vorticity-acceleration-free velocity field in space-times}, Gen. Relativ. Gravit. {\bf 51}:127 (2010).
 %
 \bibitem{Ramos}
M.~P.~M.~Ramos, E.~G.~R.~Vaz and J.~Carot, \textit{Double warped space-times}, J. Math. Phys. {\bf 44} (10), 4839
(2003). 
%
\bibitem{Rani03}
 R.~Rani, B.~Edgar and A.~Barnes, \textit{Killing tensors and conformal Killing tensors from conformal Killing vectors}, Class. Quantum Grav. {\bf 20} (11), 1929 (2003). 
%
 \bibitem{Sharma10}
R.~Sharma and A.~Ghosh, 
\textit{Perfect fluid space-times whose energy-momentum tensor is conformal Killing}, J. Math. Phys. {\bf 51},  022504
(2010). 
%
\bibitem{Wang20}
B.~Wang, J.-Z.~Qi, J.-F.~Zhang, and X.~Zhang, \textit{Cosmological model-independent constraints on spatial curvature from strong gravitational lensing and SN Ia observations}, Astrophys. J. {\bf 898}:100 (2020).
%
\bibitem{Melia22}
 J.-J.~Wei and F.~Melia, \textit{Exploring the Hubble Tension and Spatial Curvature from the Ages of Old Astrophysical Objects}, Astrophys. J. {\bf 928}:165 (2022). 
%
\end{thebibliography}
\end{document}